# A 6.3-Nanowatt-per-Channel 96-Channel Neural Spike Processor for a Movement-Intention-Decoding Brain-Computer-Interface Implant

Zhewei Jiang, Jiangyi Li, Pavan K. Chundi, Sung Justin Kim, Minhao Yang,
Joonseong Kang, Seungchul Jung, Sang Joon Kim, and Mingoo Seok

*Abstract*—This paper presents microwatt end-to-end neural signal processing hardware for deployment-stage real-time upper-limb movement intent decoding. This module features intercellular spike detection, sorting, and decoding operations for a 96-channel prosthetic implant. We design the algorithms for those operations to achieve minimal computation complexity while matching or advancing the accuracy of state-of-art Brain-Computer-Interface sorting and movement decoding. Based on those algorithms, we devise the architect of the neural signal processing hardware with the focus on hardware reuse and event-driven operation. The design achieves among the highest levels of integration, reducing wireless data rate by more than four orders of magnitude. The chip prototype in a 180-nm high-$V_{TH}$, achieving the lowest power dissipation of 0.61μW for 96 channels, 21× lower than the prior art at a comparable/better accuracy even with integration of kinematic state estimation computation.

## I. INTRODUCTION

ADVANCES in brain-computer-interface (BCI) research is aiding the development of prosthesis for patients with limited mobility. Prosthesis can be categorized as passive or active. While passive prosthesis only provides structural support for patients, active prosthesis can perform the patients' intended motor function, autonomously or controlled. BCI can aid active prosthesis by mapping their neural activities to the intended movements and actuating them. Hence, BCI systems are invaluable in limited mobility rehabilitative services [1-9].

A prosthetic BCI operates by measuring neural activity and inferring the intended movement based on a learned model (cortical map) that relates the neural behavior to the movement intention [10, 11]. Neural activities useful for motor intention decoding can be sampled directly from residual muscle activation near the prosthesis site [12-14] or deep within the motor cortex in the brain [1-9]. Any neural signals encoding motor intention can support BCI prosthesis if the encoding scheme can be reliably modeled. For locked-in patients without residual muscle activation, only central nervous signals can aid the prosthesis. Extracellular spiking activity from pre-motor or motor cortex is currently the state of the arts for upper limb movement decoding [7], outperforming non-invasive systems based on signals such as electroencephalogram (EEG) [8] or magnetoencephalography (MEG) [9]. Unlike EEG or MEG, the detection of spiking neural signal requires surgical procedure to implant probes and peripheral support devices. The invasiveness places additional physical design constraint, e.g. power limit of 40mW/cm$^2$ [15], to the already complex algorithmic challenges.

In this paper, we present a full stack design and the prototype of a prosthetic *Neural Spike Processor* (NSP), to be placed in between sensor front end and wireless neural signal communication. The NSP decodes neural spike information into direction and velocity of intended muscle movement, enabling rehabilitative services for patients. We modify and improve existing neural decoding algorithms to implement our design.

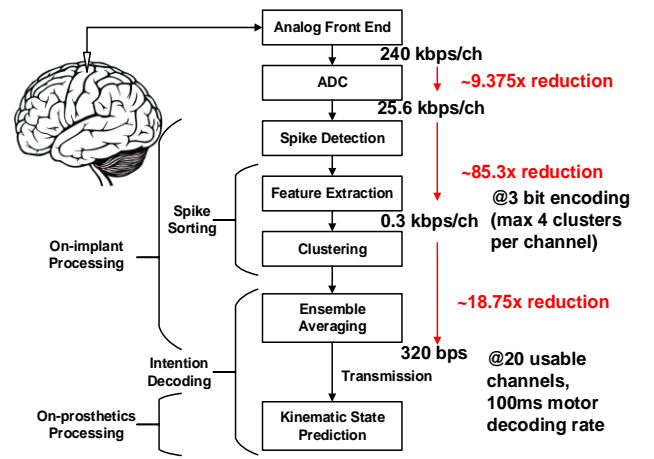

Fig.1. Proposed prosthetic BCI task flow

The remainder of the paper is organized as follows. In Section II, we first briefly present the scope and tasks involved in the BCI system. In Section III, we detail our spike sorting algorithm. In Section IV, we present our neural decoding approach and evaluate the accuracy and cost of our hardware. In Section V, we present the architecture of the NSP chip. In Section VI, we present the prototype and measurements. Finally, Section VI concludes the paper.

## II. SYSTEM OVERVIEW

Fig. 1 shows the processing stages of the targeted extracellular spike BCI system. The first stage is an implanted electrode array. The electrode array senses extracellular potentials, which originate from surrounding neurons. The signals are then filtered with band-pass or low-pass filters. Following the filtering stage, analog-to-digital converters (ADC) digitize the extracellular voltage signals and produce multi-channel digital

This work was supported by National Science Foundation, Catalyst Foundation, Samsung Electronics, and Wei Family Private Foundation. (*Corresponding author: Mingoo Seok*)

Zhewei Jiang, Jiangyi Li, Pavan K. Chundi, Sung Kim, Minhao Yang, and Mingoo Seok are with the Electrical Engineering, Columbia University, New York, NY 10027 USA (e-mail: ms4415@columbia.edu). Joonseong Kang, Seungchul Jung, Sang Joon Kim are with Samsung Electronics, Suwon, Republic of Korea.

data streams for later stages of the BCI system. The NSP initiates after the ADC stage of the BCI and terminates before the transmission of partially decoded neural data off chip.

The motor intention-decoding task is delegated to both on-chip hardware processing at the implant site (NSP) and off-chip software processing at the prosthesis site. The on-implant computation includes spike detection, spike sorting (which includes feature extraction and clustering) [16, 17], and partial computation of intention decoding which estimate the movement state via ensemble-regressed spiking events. The near-prosthesis computation concludes the rest of intention decoding, notably the Kalman filter (KF) operation, which finalize the kinematic state prediction.

The main physical constraints of an implantable BCI device is power efficiency. The temperature sensitivity of the implant site tissues can render the targeted neurons useless when exposed to high power implant. Among the tasks performed by the implant, data transmission from implant to prosthesis is the dominate power consumer for existing BCI implants, amounting to 17mW for this design assuming radio transmission at 750pJ/bit [18, 19]. Hence, our NSP design is driven by the objective of achieving the highest data rate reduction while performing the minimum computation on implant.

Based on the prior works [5-6, 19-21], we optimize and improve the algorithms for spike sorting and intention decoding to reduce on-chip computational complexity while improving the decoding accuracy. Consider a typical 96-electrode array sensing at 8-bit resolution at 30 kHz for prosthetic BCI, the data rate is nearly 3MB/s without any on-implant processing (Fig. 1). If the full data streams are transmitted off-chip entirely, the required power would make the system unsuitable for long term deployment. Therefore, we implement spike detection, feature extraction, sorting, and decode estimation on the implant, reducing the wireless data rate by more than *four* orders of magnitude.

The functions of the NSP is as follows. The processor receives the 96 channels of streams and detects neural activations (spikes) by thresholding the action potential. The NSP then performs spike sorting by their waveform shape. The sorted neurons are then analyzed for their spiking behavior. The neurons' efficacy in kinematic information encoding determines their usage in the following stages. The motor intention decoding uses a cortical model to map instantaneous spiking rates from the selected neuron population to kinematics. Instantaneous spiking rates are in practice spike counts in time bins, which are typically around 100 ms. The cortical map is attained by regressing training data, typically at the same step when we identify the neurons that encode significant kinematics information. Finally, we adopt Kalman filter (KF) for decoding but make a modification on how the predictor is computed. In the proposed method, instead of each neuron making its own kinematic prediction, the ensemble of selected neurons makes one prediction based on prior regression results. The rationale of our modification is in the computing cost as well as the high Poisson noise of individual neuron's spiking rate.

This calibration procedure requires in-patient experiments, and is typically very hard to fully automate in the deployment system. The typical calibration process is as following. A lock-in patient with a BCI implant is asked to imagine a preprogramed movement while the BCI system asserts a small amount of the control over the visual feedback of said movement (screen or prosthesis). Gradually, the BCI's decoded results are asserted more and more influence over the preprograming of prosthesis until the movement is entirely driven by the BCI system. The cortical map developed by outpatient analysis is used to make the initial motor prediction. The rigorous training of the BCI decoding features relies heavily on the neural plasticity of the patient's neurons [3, 22]. The calibration step is much more of a rehabilitative reconfiguration process on the motor cortex neurons than a setup step on an out-of-the-box working machine. While the decoding success is dependent on the patient's neural plastic health, the initial decoding input greatly affects the viability of the BCI system.

### III. SPIKE SORTING

Each extracellular electrode measures the activity of neurons in its proximity; hence, a sorting process is required to differentiate spikes by their originating neurons. The basis for sorting is the spike waveform's shape, under the assumption that multiple neurons are at various distances to the electrode and they have unique internal ionic gating states. The varying distances through the brain tissues to the electrode have varying filtering effect; the ionic gating states affect the spike shape such as relaxation and pre-spiking depreciation. The spike sorting process uses selected waveform markers to identify the individual neurons. Spike sorting provides substantial savings in power. Assuming 3-bit identifier for 32-sample 8-bit resolution spike waveform, spike sorting can achieve 85X reduction in data transmission (Fig. 1). Our design goal concerning spike sorting is to implement algorithm and hardware that consumes minimal energy and area at high accuracy.

*A. Prior Work: Constrained Bayesian Boundary Sorting*

We will briefly describe the low-power unsupervised spike sorting in this subsection, our earlier work [20], the proposed NSP sorting algorithm is based on. We categorize prior works on on-chip spike sorting by their decision metric, namely distance to template [19,21,23] or boundary [20]. Due to the high area and power cost of a multiplier, all of the distance-based sorters use L1 distance metric (the sum of absolute difference of two vectors) on time-domain features.

Here, we aim to perform the spike sorting in a per-channel basis. The features used in the spike-sorting task are the maximum (peak) and minimum (trough) value of each spike waveform. Using simple time-domain features is not optimal for accuracy, but is a trade-off in favor of lower computational cost. The peak and trough values of each action potential waveform construct the feature space of the clustering problem. Given the 2-dimensional feature space, the theoretically optimal sorting solution is a set of *free-form*

Bayesian boundaries from clusters' distribution intersects. Any improvement in accuracy can only be derived from better feature selection/extraction. This approach is, however, far too costly to on-implant processing as it requires large memory and complex computations. Instead, we project two of single-feature Bayesian distribution boundaries to the 2-feature space, forming a set of grid-constrained boundaries.

The Fig. 2 shows the detailed unsupervised and online learning process of the *constrained Bayesian boundary sorting* algorithm. The first step is to identify boundaries in the feature space. For each detected spike (Fig. 2 top left), it updates the two histograms of the peak and trough values (Fig. 2 bottom left). After the specified number of spikes are used for constructing histograms, the local minima distributions are stored as Bayesian boundaries (Fig. 2 bottom right). These boundaries orthogonally partition the feature space, with each partition identifiable by a pair of indexes (Fig. 2 top right).

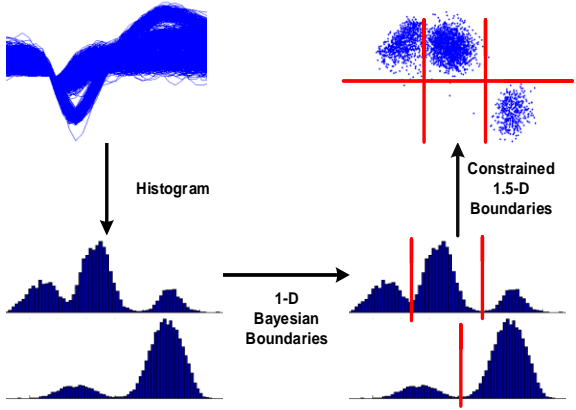

Fig. 2. Constrained Bayesian boundaries found in an unsupervised fashion

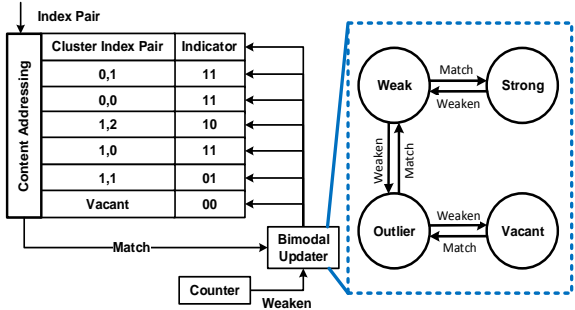

Fig. 3. CAM for cluster status and update rule

After the boundaries are found, each feature space partition is to be assigned a status as either a cluster or unnecessary segmentation. This is done by updating the confidence level of the cluster status of each partition in the feature space. The specific steps are as follow. First, the features of an incoming spike is compared with the stored boundaries. This locates the specific partition that the spike belongs in. The pair of indexes is then searched in a CAM (content addressable memory) (Fig. 3) in which data matched in a single operation. Associated with each partition is a status is a 2b indicator (00-vacant, 01-outlier, 10-weak cluster, 11-strong cluster) as cluster status. If there is a hit in the CAM, its confidence level is increased. The controller periodically decreases all entries' indicators once per N spikes to remove outliers.

After a user-specified amount of training, the post-training spike-sorting process is deployed. This process performs the same computation for finding an index pair as the training process, but it no longer updates the CAM indicators. A min function is performed on the CAM results (Fig. 4). The spike is then assigned to the closest partition that is a valid cluster.

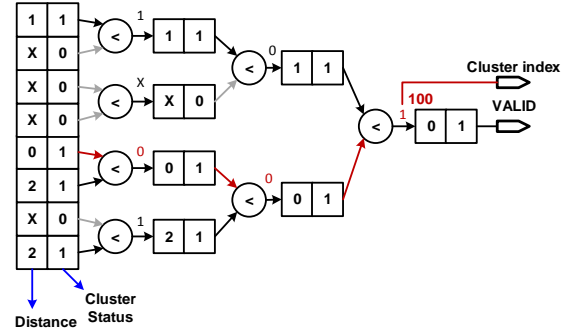

Fig. 4. Decision tree branch identification by finding boundaries closest to input in the feature space.

The algorithm is similar to a decision tree. The branching conditions are Bayesian boundaries. The cluster status update serves as a pruning process on the decision tree.

### B. Supervised Training of the Constrained Bayesian Boundary Sorting

For the NSP development, we designed a similar spike sorter based on the constrained Bayesian boundary model yet uses the offline-training model for improving accuracy. The accuracy of the online-trained model is weaker for several reasons. One factor is the suboptimal feature selection. Another factor is that decision tree pruning can make part of the feature space unreachable. Some of this problem is recovered by the adjacency checking of the feature to the known branching nodes. Still, it is not fully reliable since not all nodes are Bayesian optimal for that modality. Furthermore, as constant retraining is required for intention decoding, offline trained sorting algorithm would not incur additional retraining that is not tied to decoding retraining while providing higher accuracy overall.

Our *supervised-trained* sorter uses two voltage samples directly selected from the waveform, but not necessarily the peak and trough values as used in the unsupervised training version. Given pre-activation and relaxation behavior of neurons, peak/trough are often not accuracy-optimal time domain samples for clustering. To improve accuracy while keeping the dimensionality low, we sweep features during offline training to find the two best performing features. We perform a parameter sweep to find the best performing pair of indexes, across all the spike waveforms used in the training process, as the decision tree's feature space.

We construct the decision tree in offline training (Fig. 5). First, we estimate the distribution of data points with Gaussian kernels. We then sweep for boundaries under the orthogonality constraint. The boundary information (direction, order, and values) is stored in implants. We limit the maximum number of neurons per channel to four based on the observation of our target datasets [28] where the maximum number is four. In the case of more than four clusters present in a channel, the decoding performance is not affected with a heuristic that no more than three neurons from one channel are selected for intention decoding.

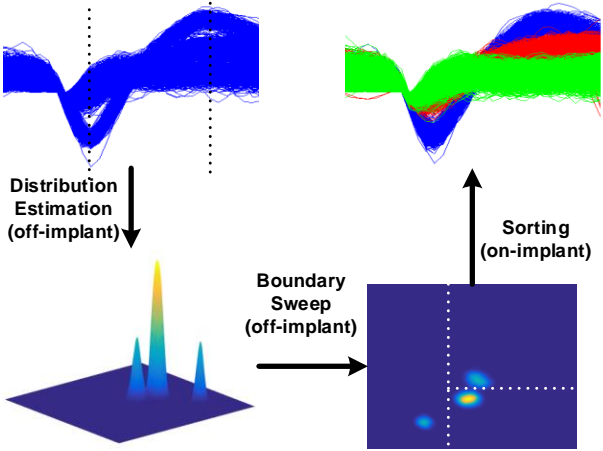

Fig. 5. Visuallization of spike soting via Baysian approximate decision tree.

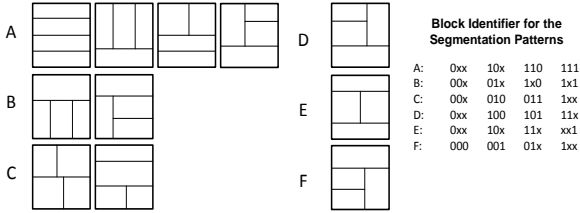

Fig. 6. Feature space segmented block identification coding.

Under the constraint limiting to four clusters per channel (3 boundaries, ordered), eleven unique segmentation patterns exist in the feature space (Fig. 6). Here, the diagonally symmetric patterns are considered a single class. Each spike, as a point in the space, can be located by simple comparison with the segmentation boundaries (these are the decision tree branching conditions). Its cluster is identified by the 3-bit comparison results. The comparisons against each of the boundary, combing with the feature organization, produce a unique marker that identifies it as a cluster within a channel.

The 11 segmentation patterns are organized into six groups. For all patterns in the same group, the four blocks share the same boundary comparison identifier. This also means the order of the features being used for of decision tree branching condition are in the same order. This allows an encoding scheme that reduces memory needed to characterize the channel as compared to the online-trained implementation in [21]. Online training require storage for the full 4x4 grid since the optimal segmentation pattern is unknown. This requires three boundaries along each dimension, and a large hash table to store 16 blocks' information (cluster validness, its associated weight for intention decoding). The proposed offline variant thus has more than 50% reduction in memory.

### C. Accuracy and Cost Evaluations

We use the data measured from mice [24] to evaluate the sorting accuracy of our proposed algorithm. The dataset (D1, D2, D3, and D4) contains channels having signals from two to four neurons. The sampling rate is 40 kHz, band-pass filtered between 300 Hz and 5 kHz.

To compare the proposed constrained Bayesian boundary sorting to the L1 norm, we use bit-accurate hardware simulation to derive the result which shows no degradation from the ideal algorithmic baseline. In Table I, we show that the boundary-based sorting has a comparable accuracy with the L1 norm, both using two features. The accuracies vary for the test datasets as they are of different SNR and spike shape similarity.

The decision boundary based on L1 norm is not constrained to orthogonal grids in the feature space, which can theoretically outperform the proposed constrained boundary model. However, as the features are voltage samples taken directly from the spike waveform, its non-idealities (noise) account for its sorting performance. Since spike's peak, trough, and relaxation slope are primarily driven by different ion pumps, different parts of a spike waveform have different variances. L1 metric does not consider this phenomenon; thus, its sorting accuracy is negatively affected.

TABLE I. SPIKE SORTING ACCURACY

| Decision Metric | Datasets (Number of neurons) | | | |
|---|---|---|---|---|
| | *D1(2)* | *D2(3)* | *D3(4)* | *D4(4)* |
| L1 Norm | 99.97% | 99.25% | 91.39% | 89.00% |
| Const. Boundary | 99.99% | 99.19% | 91.61% | 89.49% |

TABLE II. COMPARISON OF MEMORY AND COMPUTING COMPLEXITY IN L1 NORM AND CONST. BOUNDARY

| | *L1 Norm* | *Const. Boundary* |
|---|---|---|
| Memory/channal | 8B | 3B + 4b |
| Computational Complexity | 12(+/-) + 3(<) | 3(<) + 1(CAM) |

The advantage of the decision tree is that it requires much less on-chip memory and computation than the distance-based technique as shown in Table II. The memory required in the presented sorter per channel is 3 bytes for storing boundary information, 4 bits for segmentation pattern (total 28 bits) for $n$-cluster per channel ($n < 5$). By comparison, a 2-feature L1 metric requires $2 \cdot n$ bytes (64 bits when $n=4$). In term of the computation complexity, the 2-feature L1 requires $3 \cdot n$ 8-b additions/subtractions and $n$-1 8-b comparisons, while our proposed model only needs three 8-b comparisons and one 3-bit 4-entry CAM read.

## IV. INTENTION DECODING

### A. Theoretical Basis

The standard Kalman filter (hereafter referred to as Kalman filter) has been one of the state-of-art decoding methods in cortical spiking-based motor intention BCI. Its observation

(state) model is based on the preferred direction property of select neurons. The observation-based state estimate, i.e. instantaneous velocity, is modeled as a weighted sum of target neurons' time-binned spiking rate. The weight is derived solely from error covariance.

The Kalman Filter's weighting basis does not regard the neuron's firing behavior across intended movement directions, using the same error variance in all cases. This assumes that a neuron's spiking rates maintains the same signal-to-noise ratio regardless if the intended movement is in its preferred direction. In actuality, the spiking rate variance during a state of excitation, inhibition, and idle are not constant, thus the linear regression of single neuron observation has uneven reliability depending on the a priori estimate. Furthermore, even if the spiking variance is always constant, high firing rate (preferred direction) is more reliable simply due to higher SNR.

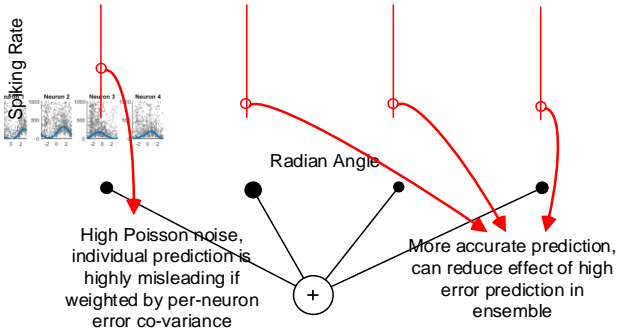

Fig. 7. Multivariat regression of ensemble neurons reduce the effect of noise

The more biologically realistic model has been validated by some studies that have fitted the excitation response in cosine or wrap-around Gaussian functions [25]. This is the tuning curve of the neuron. If the tuning curve model is used to improve the Kalman filter, it will take the form of a separate iterative process to modify the error covariance from the previous step to one that corresponds the current state transition prediction. This will require a large lookup table for the error covariance values from training experiments and would increase the cost of computation. Furthermore, neural models are non-stationary. Changes in neural behavior would require more drastic correction to the tuning curves if an adaptive model is needed [26].

We propose Ensemble Observation Kalman filter (EOKF) to adapt an entirely different approach to weighting neurons' predictions. By reducing the probability distributions of many individual observations to a single population ensemble observation, the EOKF uses a population vector model and determines the weights through multi-variate regression. Instead of the individual spiking rate modulated only by error covariance, it is further modulated by its own excitatory state, benefiting from the signal of a higher SNR. As shown in Fig. 7, an example of 4 neurons firing when the movement is in direction of radian angle of -1. For neuron 1, the intended movement does not correspond to excitation. However, the baseline firing is noisy and still has a high spiking rate. For neuron 2 to 4, the spiking rate is close to the tuning curve profile. The baseline Poisson noise dominated spiking from neuron 1 is thus compensated in an ensemble, since the weights acquired via multiple regression account for the behavior of excited neurons at the same time. Hence the weighted vector sum is a better observation than the single neurons'.

B. *Ensemble Observation Kalman Filter*

In this subsection, EOKF is presented in detail. This filter not only improve decoding accuracy, but also reduce on-chip computation workload, and minimize the data rate in transmission to off implant.

The standard KF for motor intention decoding is:

$$x_{k+1} = A\,x_k + w_k \qquad (1)$$

$$z_k = H\,x_k + q_k \qquad (2)$$

The term $x$ is the state, i.e. position, velocity, etc. $A$ is the state transition matrix; $w$ is the state noise (typically zero-mean Gaussian variable); $z$ is observation, i.e., binned spiking rates; $H$ is the observation matrix, i.e. cortical map; $q$ is observation noise (zero mean Gaussian variable); subscript $k$ is the time step (100ms, determined from parametric sweep).

Eq. (1) describes the state transition in a Markov chain. It simply provides a movement constraint from one moment to another. The constraint acts as a dampener to avoid overly aggressive prosthetic movement. Eq. (2) describes the more interesting behavior as it formulates the cortical mapping between kinematic state and neural activity. Given an estimated state, Eq. (2) reconstructs the expected spiking rates from the selected population.

The standard KF updates its state and parameters iteratively, beginning with an a priori estimate of next kinematic state via the Markov process. A priori estimate is given as:

$$x^-_k = A\,x^-_{k-1} \qquad (3)$$

The standard KF's posterior estimate combines the passive estimate from state transition and the weighted error between expected observation and actual observation:

$$x_k = x^-_k + K_k(z_k - H\,x^-_k) \qquad (4)$$

The error weights are known as Kalman gain $K$, computed as:

$$K_k = P^-_k H^T (H\,P^-_k H^T + Q)^{-1} \qquad (5)$$

The Kalman gain is updated with the error covariance matrix P and the measurement error matrix Q. Eq. (5) represents a substantial computation load, since the number of neurons selected for motor intention decoding is relatively large, typically between 20 to 50 [7, 25], making $H\,P^-_k H^T$ at least a 20×20 matrix. In particular, there exist no closed-form solution to inverse such large matrix. Employing a numerical method for this inversion problem inevitably increases power and area overhead for an implant.

To reduce this computational complexity, we propose an inverse form of observation-to-state transition as a committee machine, essentially changing (2) to:

$$x_k = E\,z_k + q_k \qquad (6)$$

in which the expected state is constructed with ensemble spiking rates, weighted by $E$. Fig. 8 illustrates this. After this operation, the cortical map $H$ now becomes trivial (identity matrix) in EOKF, therefore reducing the computational workload henceforth. The proposed filter has the same a priori estimate as the standard KF, as shown in Eq. (4).

In the EOKF, (4) and (5) are reduced to:

$$x_k = x^-_k + K_k ( E z_k - x^-_k ) \quad (7)$$

$$K_k = P^-_k (P^-_k + Q )^{-1} \quad (8)$$

With the new smaller filter (i.e., Eqs. (4), (7), (8)), the lowest data bandwidth of the entire BCI system is located at computation of $E \cdot z_k$. This term has the equivalent data rate as the final decoder output ($x_k$). Therefore, we perform only the computation of $E \cdot z_k$ and all the other computations are offloaded to prosthetics sites where power and area budget is much greater. This partition also includes the computation of (3) in the prosthetics site since there is no data dependency.

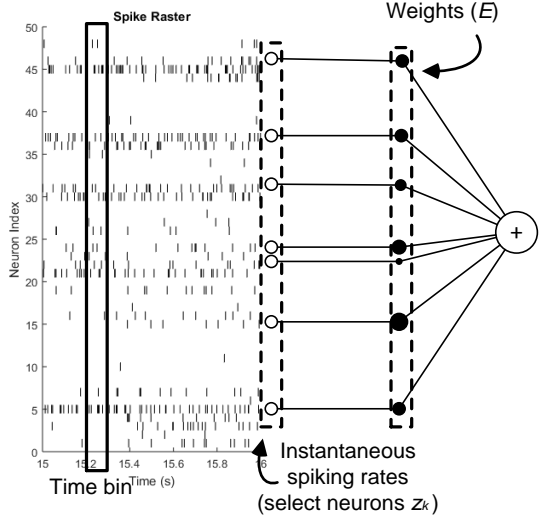

Fig. 8. Instantaneous spiking rates weighted in ensemble average for the observation based estimate of a kinematic state.

TABLE III. COMPARISON OF NUMBER OF CALCULATIONS IN EOKF AND STANDARD KF (20-NEURON CORTICAL MAP)

| Equation | Number of Calculations (Mult/Add/Div) | |
|---|---|---|
| | *Standard Kalman* | *Ensemble observation Kalman* |
| (4) | 4/2/~ | 4/2/~ |
| (5)/(7) | 80/80/~ | 46/46/~ |
| (6)/(8) | 32180/32060/1180 | 10/9/4 |
| (9) | 8/8/~ | 8/8/~ |
| (10)/(11) | 88/84/~ | 8/8/~ |

The standard KF posterior error covariance matrix updates at each time step (100ms, determined from parametric sweep).

$$P^-_k = A P_{k-1} A^T + W \quad (9)$$

$$P_k = (I - K_k H) P^-_k \quad (10)$$

The error covariance estimate is updated with state transition and state error variance $W$ in (9). The posterior error covariance incorporates the observation error in the form of Kalman gain in (10). In EOKF, (10) is reduced to (11).

$$P_k = (I - K_k) P^-_k \quad (11)$$

As mentioned, the posterior error covariance matrix width is now the number of state elements, 3 (x/y directions, and velocity), instead of the number of neurons.

Finally, in Table II, we compare the number of multiplications, additions, and divisions in the proposed EOKF and the standard KF. We assume to select 20 neurons for intention decoding. We can find 1-2 orders of magnitude reduction in the operations, even for the case of including the computations both on implants and prosthetics site.

### C. EOKF Evaluation

To evaluate the performance of our proposed EOKF, we use an upper-limb reaching data set [28] from the Database for Reaching Experiment and Models (DREAM) [29]. The task is the standard 2D equal-distance 8-target center-out-reach-and-return performed by a Rhesus monkey well trained in the experiment. Only the velocity vector of the hand movement in the x, y plane is used as kinematic state, same as the velocity-Kalman study [6]. The data set contains 194 trials of the 196-neuron spiking traces from the motor cortex.

In this study, the data is used for offline reconstruction of native movements. We train the filter state parameters (transition matrices, error variances) with 80% of the data (randomly selected for each trial). Post-training decoding (reconstruction) is done on the remaining 20%.

To evaluate the accuracy, we use bit-accurate hardware simulation to perform the kinematic state estimation component of EOKF. The proposed algorithm outperforms the standard KF with 28% lower trace reconstruction error on average, at a standard deviation of 198% and a high kurtosis of 39.4. This outperformance comes from the advantage that the proposed EOKF always has better observation than the standard KF. In the proposed EOKF, the error variance of the estimate has an upper bound that is equal to the lowest error variance of its members, in which case, the committee is trivial having the single member determine the output.

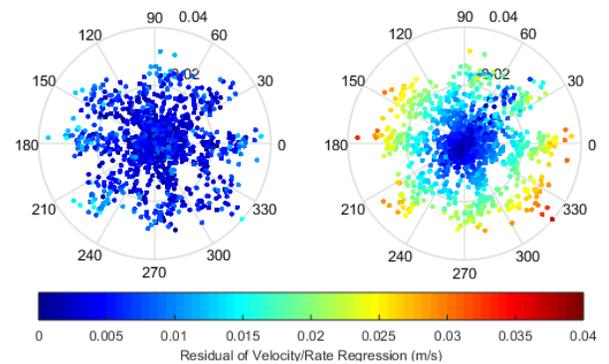

Fig. 9. State prediction error of ensemble prediction (left) and single neuron prediction (right) across angles and speed.

The advantage of the committee machine also manifests in error variance consistency. Fig. 9 demonstrates the regression residuals of an ensemble and a single neuron. Each dot represents a kinematic state instance. The residual is color-mapped to show that the accuracy of single neuron model (used in the standard KF) is less consistent across directions and movement speed than the ensemble model used in the proposed EOKF. The error variance matrices are assumed invariant in both the standard KF and EOKF; hence, the ensemble better fits the variance model due to its evenness. The same result is faithfully produced in hardware as there is no channel collision in the data stream.

## V. Processor Architecture

Based on the optimized sorting and decoding models, we prototyped the 96-channel Neural Spike Processor (NSP). The NSP architectural design focuses on resource sharing to minimize area and power. In addition, we exploit data sparsity inherent in neural spike activities to implement event-driven computation for lower resources. All stages after parallel spike detections can share hardware without a separate fast/slow clock domain, data stream multiplexers, or additional controller for time multiplexing.

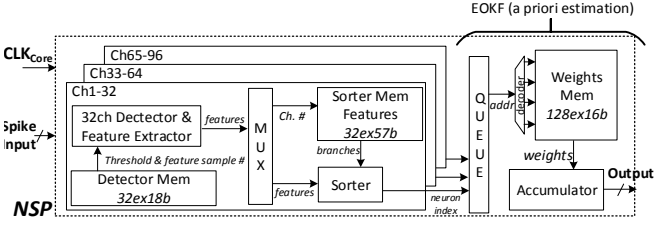

Fig. 10. The architecture of the proposed NSP.

Fig. 10 shows the on-chip hardware architecture of the proposed NSP. It includes the modules for spike detection, spike sorting, and the ensemble estimation. The NSP starts with 96 non-overlapping spike detectors each of which integrates a simple feature extractor, delivering spike features directly to the sorter modules. We grouped 32 of the spike detectors to share a single set of sorter hardware. Each sorter hardware has the memory entries of boundaries for the 32 channels and those for neuron identifiers for the data transfer of spike events to the following decoding module.

The spike waveform length mainly determines the group size, i.e., the number of spike detectors that share sorting hardware. In our design, each waveform has 32 samples. Since the waveform detection is non-overlapping, no channel can generate more than one spike event within 32 cycles. Since a single sorter can perform one sorting per cycle, 32 channels can share a single sorter without congestion.

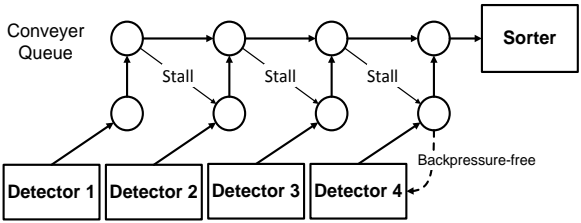

Fig. 11. Conveyer style queue of a 4-detector example

Specifically, at the event of a spike, the detector enters an event token (i.e., spike features) onto a conveyer style queue (Fig. 11). The queue entry points from detectors have a simple stalling rule that gives priority to the token already on the conveyer, the stalled token then attempts to enter the queue in the subsequent clock cycles until a free spot is available. With the non-overlapping detection and the defined samples per spike, token stalling does not generate backpressure to earlier stages. The conveyer queue does not preserve the time order of spikes. However, the order is irrelevant downstream where only spiking rate is considered. Thus, this minimally affects the accuracy of the rate based decoding schemes that our EOKF belongs to. No spike is stalled beyond the time bin edges (nor would it be problematic as spike rate has relatively high Poisson noise) to affect the following rate computation.

The sorter modules, each responsible for 32 channels, consume event tokens in the order of the channel array if no collision is present in the data streams, but in cases of token stalling, channel order is compromised. As a result, a neuron index (i.e., address) must be included in the token to retrieve correct boundary information used in sorting. The sorter checks the incoming features against boundaries stored in the Bayesian boundary memory. Note that we use the coding based on fixed segmentation patterns in the feature space (Sec. II). Thus, the outputs of the sorter are the indexes (addresses) for the memory storing weights for decoding.

Finally, the architecture has a single ensemble-regression module (memory and accumulator). This decoding module does not require resource sharing since all data paths converge to a single register per state element in $Ez_k$. Instead of multiplying spike rates by weights ($E$), we perform memory read for retrieving a coefficient and then accumulate it using a single adder upon every spike event. This is equivalent to multiply-accumulate since the instantaneous spiking rate is a count of spikes in a fixed-duration time bin. This architecture can reduce silicon area and thus leakage power.

A typical challenge of event-driven implementation in place of a scheduled one is potential data collision hazards. In our architecture, we can have collisions among the three spike sorting modules when they try to access the ensemble-regression module at the same time. Unlike spike detection, the sorter has no hardware constraint for token generation. With a finite amount of buffer memory, therefore, token loss is possible. Practically, however, token loss is improbable since only a small subset of all sorted neuron channels (e.g., 20 ~ 50 in typical experiments) is selected for intention decoding. In our test, no token loss or even collision occurs even with all sorted neuron channels considered valid. In the unlikely event that a token is lost, its effect is minimal since the ensemble-regression module can easily tolerate small loss.

## VI. Prototype Measurement

We prototyped the 96-channel NSP in a 0.18-μm CMOS technology. The technology is chosen since leakage power is the major energy efficiency bottleneck in the NSP [30]. Fig. 11 shows the chip die photo. The total area is only 1.86 mm$^2$. The 96 detectors, three sorters, and one EOKF decoder take the similar silicon footprints (Fig. 12).

The power consumption of the NSP is data rate dependent thanks to its event-driven operation. Fig. 13 shows the power consumption across different supply voltages (0.8, 0.5, and 0.3V) and across different input spiking rates. Operations at each step may be terminated as soon as information of each spiking event's efficacy at changing the decoding output becomes available. Spiking event from electrode channels that do not factor in the ensemble is not active at all and spiking from useful channels but not in selected ensemble is stopped after sorting. The power consumption scales with patient movement intent.

uncertainty constraints as it increases C-to-Q and logic delay. This invites us to scale supply voltage to the subthreshold level of 0.32 V (Fig. 14), the average minimum voltage that maintain NSP functionality. The 96-channel NSP consumes only 0.61 μW.

In Table III, we compare the NSP to the state-of-the-art BCI processors [20, 23, 31-33]. The proposed NSP achieves 21× smaller power dissipation than [23]. It also demonstrates the highest level of integration, namely the first end-to-end integration of neural signal processing at better accuracy over prior arts.

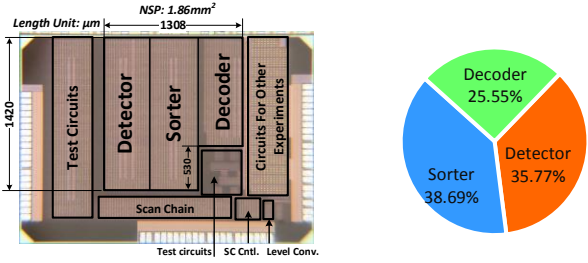

Fig. 12. Die photo and area breakdown.

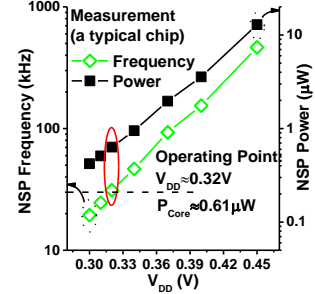

Fig. 14. Power and performance of the NSP.

## VII. CONCLUSION

In this work, we present a nanowatt neural spike processor for a movement-intention-decoding neural interface implant. We devise/optimize algorithms and architecture for hardware and energy cost. Our design provides substantial resource savings from prior arts. We verified our algorithm using data driven testing for spike sorting and intention decoding, and with additional boundedness analysis for the proposed ensemble observation model. Our proposed hardware architecture enables effective hardware sharing and event-driven architecture, thereby substantially reducing area and power dissipation. The NSP not only achieves the highest level of functional integration from spike detection to the intention decoding but also marks a record power efficiency.

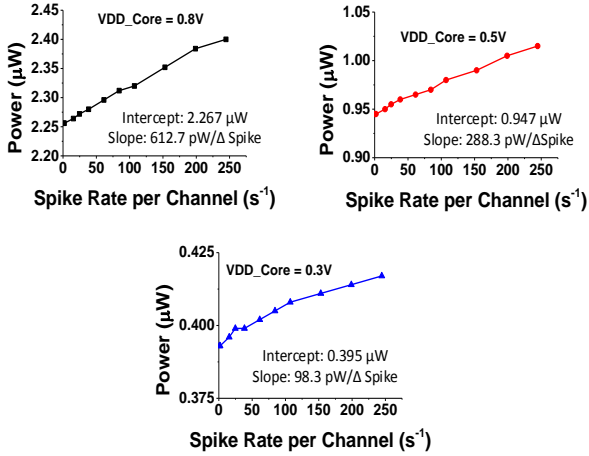

Fig. 13. Spike rate dependency of the NSP power.

The target clock frequency of the NSP is the sample frequency of the front-end sensor ADC, at 30 kHz in our system. Our detecting, feature extracting, sorting, and decoding models exhibit substantially small computational complexity and therefore can easily meet the 30-kHz timing requirement. The low frequency also helps lower the hold time

TABLE IV. COMPARISON TO THE PRIOR BCI PROCESSORS

|  | **This work** | **[6]** | **[10]** | **[17]** | **[18]** | **[19]** |
|---|---|---|---|---|---|---|
| **Process (nm)** | 180 | 65 | 65 | 65 | 130 (sim) | 180 |
| **No. of Channels** | 96 | 96 | 128 | 16 | 32 | 64 |
| **Detection Algorithm** | AT* | AT | ICD* | AT | NEO* | NEO |
| **Sorting Training** | Supervised | Unsupervised | Supervised | Unsupervised | Unsupervised | Unsupervised |
| **Sorting Algorithm** | Decision Tree | Bayesian | K-Means | O-Sort | Gap Stat K-Means | C-Sort |
| **Spike Dataset** | [8] (Recoding) | [8] (Recoding) | [20] (sim) | [21] (sim) | [22] (Recoding) | [20] (sim) |
| **Sorting Accuracy** | 89~99% | 95% | 77%~87% | 75% | 91% | 67%~93% |
| **Decoding** | **Y (partial)** | N | N | N | N | N |
| **Core $V_{DD}$ (V)** | 0.32 | 0.6 | 0.54 | **0.27** | 1.2 | 1.8 |
| **Core Power/Ch. (nW)** | **6.3** | 1740 | 175 | 108.8 | 750 | 2313 |
| **Core Area/Ch. (mm²)** | 0.0194 | 0.12 | **0.003** | 0.07 | 0.023 | 0.094 |

*AT = Absolute Thresholding, ICD = Integer Coefficient Detector, NEO = Non-linear Energy Operator

## REFERENCES

[1] Wu, W., Gao, Y., Bienenstock, E., Donoghue, J. P., & Black, M. J., 2006. Bayesian population decoding of motor cortical activity using a Kalman filter. *Neural computation*, 18(1), 80-118.

[2] Schwartz, A. B., Cui, X. T., Weber, D. J., & Moran, D. W., 2006. Brain-controlled interfaces: movement restoration with neural prosthetics. *Neuron*, 52(1), 205-220.

[3] Guggenmos, D. J., Azin, M., Barbay, S., Mahnken, J. D., Dunham, C., Mohseni, P., & Nudo, R. J., 2013. Restoration of function after brain damage using a neural prosthesis. *Proceedings of the National Academy of Sciences*, 110(52), 21177-21182.

[4] Jackson, A., & Zimmermann, J. B., 2012. Neural interfaces for the brain and spinal cord—restoring motor function. *Nature Reviews Neurology*, 8(12), 690.

[5] Serruya, M., Shaikhouni, A. and Donoghue, J.P., 2003. Neural decoding of cursor motion using a Kalman filter. In *Advances in Neural Information Processing Systems 15: Proceedings of the 2002 Conference* (Vol. 15, p. 133). MIT Press.

[6] Kim, S.P., Simeral, J.D., Hochberg, L.R., Donoghue, J.P. and Black, M.J., 2008. Neural control of computer cursor velocity by decoding motor cortical spiking activity in humans with tetraplegia. *Journal of neural engineering*, 5(4), p.455.

[7] Gilja, V., Nuyujukian, P., Chestek, C.A., Cunningham, J.P., Byron, M.Y., Fan, J.M., Churchland, M.M., Kaufman, M.T., Kao, J.C., Ryu, S.I. and Shenoy, K.V., 2012. A high-performance neural prosthesis enabled by control algorithm design. *Nature neuroscience*, 15(12), pp.1752-1757.

[8] Pineda, J.A., Allison, B.Z. and Vankov, A., 2000. The effects of self-movement, observation, and imagination on μ rhythms and readiness potentials (RP's): toward a brain-computer interface (BCI). *IEEE Transactions on Rehabilitation Engineering*, 8(2), pp.219-222.

[9] Mellinger, J., Schalk, G., Braun, C., Preissl, H., Rosenstiel, W., Birbaumer, N. and Kübler, A., 2007. An MEG-based brain–computer interface (BCI). *Neuroimage*, 36(3), pp.581-593.

[10] Schwartz, A. B., 2004. Cortical neural prosthetics. *Annu. Rev. Neurosci.*, 27, 487-507.

[11] Ganguly, K., & Carmena, J. M., 2009. Emergence of a stable cortical map for neuroprosthetic control. *PLoS Biol*, 7(7), e1000153.

[12] Zardoshti-Kermani, M., Wheeler, B. C., Badie, K., & Hashemi, R. M., 1995. EMG feature evaluation for movement control of upper extremity prostheses. *IEEE Transactions on Rehabilitation Engineering*, 3(4), 324-333.

[13] Ajiboye, A. B., & Weir, R. F., 2005. A heuristic fuzzy logic approach to EMG pattern recognition for multifunctional prosthesis control. *IEEE Transactions on Neural Systems and Rehabilitation Engineering*, 13(3), 280-291.

[14] Cipriani, C., Zaccone, F., Micera, S., & Carrozza, M. C., 2008. On the shared control of an EMG-controlled prosthetic hand: analysis of user–prosthesis interaction. *IEEE Transactions on Robotics*, 24(1), 170-184.

[15] Wolf, P. D., & Reichert, W. M. 2008. Thermal considerations for the design of an implanted cortical brain–machine interface (BMI). *Indwelling Neural Implants: Strategies for Contending with the In Vivo Environment*, 33-38.

[16] Rey, H. G., Pedreira, C., & Quiroga, R. Q., 2015. Past, present and future of spike sorting techniques. *Brain research bulletin*, 119, 106-117.

[17] Lewicki, M. S., 1998. A review of methods for spike sorting: the detection and classification of neural action potentials. *Network: Computation in Neural Systems*, 9(4), R53-R78.

[18] Agarwal, K., Jegadeesan, R., Guo, Y. X., & Thakor, N. V., 2017. Wireless power transfer strategies for implantable bioelectronics. *IEEE reviews in biomedical engineering*, 10, 136-161.

[19] Karkare, V., Gibson, S. and Marković, D., 2013. A 75-μw, 16-channel neural spike-sorting processor with unsupervised clustering. *IEEE Journal of Solid-State Circuits*, 48(9), pp.2230-2238.

[20] Jiang, Z., Cerqueira, J., Kim, J., Wang, Q., Seok, M., 2016. 1.74 - μ W/ch, 95.3% - accurate Spike sorting hardware based on Bayesian decision. *Symposium on VLSI Circuits* (SOVC), pp. 34-35.

[21] Jiang, Z., Wang, Q. and Seok, M., 2015, June. A low power unsupervised spike sorting accelerator insensitive to clustering initialization in sub-optimal feature space. In *Proceedings of the 52nd Annual Design Automation Conference* (p. 174). ACM.

[22] Fallon, J. B., Irvine, D. R., & Shepherd, R. K., 2009. Neural prostheses and brain plasticity. *Journal of Neural Engineering*, 6(6), 065008.

[23] Zeinolabedin, SMA., Do, AT., Jeon, D., Sylvester, D., and Kim, T., 2016. A 128-Channel Spike Sorting Processor Featuring 0.175 μW and 0.0033 mm2 per Channel in 65-nm CMOS, *Symposium on VLSI Circuits* (SOVC), pp. 32-33.

[24] Wang, Q., Webber, R.M. and Stanley, G.B., 2010. Thalamic synchrony and the adaptive gating of information flow to cortex. *Nature neuroscience*, 13(12), pp.1534-1541.

[25] Hauschild, M., Mulliken, G.H., Fineman, I., Loeb, G.E. and Andersen, R.A., 2012. Cognitive signals for brain–machine interfaces in posterior parietal cortex include continuous 3D trajectory commands. *Proceedings of the National Academy of Sciences*, 109(42), pp.17075-17080.

[26] Waldert, S., Pistohl, T., Braun, C., Ball, T., Aertsen, A. and Mehring, C., 2009. A review on directional information in neural signals for brain-machine interfaces. *Journal of Physiology-Paris*, 103(3-5), pp.244-254.

[27] Zhang, Y., & Chase, S. M. 2013. A stabilized dual Kalman filter for adaptive tracking of brain-computer interface decoding parameters. *In 2013 35th Annual International Conference of the IEEE Engineering in Medicine and Biology Society (EMBC)* (pp. 7100-7103). IEEE.

[28] Stevenson, I.H., Cherian, A., London, B.M., Sachs, N.A., Lindberg, E., Reimer, J., Slutzky, M.W., Hatsopoulos, N.G., Miller, L.E. and Kording, K.P., 2011. Statistical assessment of the stability of neural movement representations. *Journal of neurophysiology*, 106(2), pp.764-774.

[29] Walker, B. and Kording, K., 2013. The database for reaching experiments and models. *PloS one*, 8(11), p.e78747.

[30] Li, J., Chundi, P.K., Kim, S., Jiang, Z., Yang, M., Kang, J., Jung, S., Kim, S.J. and Seok, M., 2018, September. A 0.78-μW 96-Ch. Deep Sub-Vt Neural Spike Processor Integrated with a Nanowatt Power Management Unit. In *ESSCIRC 2018-IEEE 44th European Solid State Circuits Conference (ESSCIRC)* (pp. 154-157). IEEE.

[31] Karkare, V., Gibson, S., & Marković, D. 2013. A 75-μw, 16-channel neural spike-sorting processor with unsupervised clustering. *IEEE Journal of Solid-State Circuits*, 48(9), 2230-2238.

[32] Yang, Y., Boling, S., & Mason, A. J. 2017. A hardware-efficient scalable spike sorting neural signal processor module for implantable high-channel-count brain machine interfaces. *IEEE transactions on biomedical circuits and systems*, 11(4), 743-754.

[33] Zamani, M., Jiang, D., & Demosthenous, A. (2018). An Adaptive Neural Spike Processor With Embedded Active Learning for Improved Unsupervised Sorting Accuracy. *IEEE transactions on biomedical circuits and systems*, 12(3), 665-676.

[34] Quiroga, R. Q., Nadasdy, Z., & Ben-Shaul, Y. (2004). Unsupervised spike detection and sorting with wavelets and superparamagnetic clustering. *Neural computation*, 16(8), 1661-1687.

[35] Gibson, S., Judy, J. W., & Markovic, D. (2010). Technology-aware algorithm design for neural spike detection, feature extraction, and dimensionality reduction. *IEEE transactions on neural systems and rehabilitation engineering*, 18(5), 469-478.

[36] Mizuseki, K., Sirota, A., Pastalkova, E., & Buzsáki, G. (2009). Theta oscillations provide temporal windows for local circuit computation in the entorhinal-hippocampal loop. *Neuron*, 64(2), 267-280.

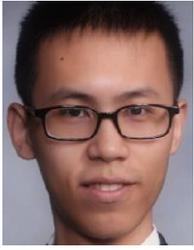

**Zhewei Jiang** received the duo B.S. degree in physics from Adelphi University, NY, USA, and from electrical engineering from Columbia University, NY, USA, in 2013 and the M.S. degree in electrical engineering from Columbia University, in 2015. He is currently pursuing the Ph.D. degree in electrical engineering at Columbia University.

He is a research assistant at VLSI Lab at Columbia University since 2015. His research interest includes neuromorphic computing architecture, neural signal processing, in-memory computation for machine learning, and other algorithm implementations.

Mr. Jiang's awards and honors include Wei Family Private Foundation Fellowship and William L. Everitt Award.

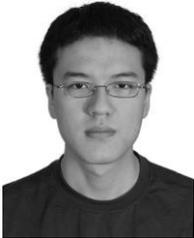

**Jiangyi Li** (S'17) received B.S. degree in Electronic Engineering from Tsinghua University, Beijing, China in June, 2012 and M.S. degree in 2014 and Ph.D. degree in 2018, both in Electrical Engineering from Columbia University. He is currently with Apple. His current research interests is low-power high-performance VLSI circuit designs including energy-harvesting systems, nanowatt power management IC, and physically unclonable function circuits.

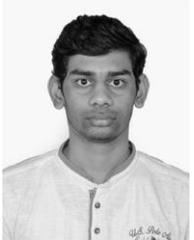

**Pavan Kumar Chundi** received B.Tech. degree from the Indian Institute of Technology Delhi, India in 2016. He obtained M.S. in Electrical Engineering in 2017 and is currently working toward Ph.D. at Columbia University, New York, NY, USA. His current research interests include accelerators for neural networks, brain machine interface circuits design and emerging memory technology.

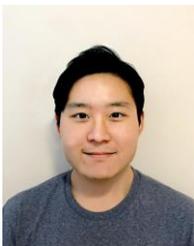

**Sung Justin Kim** (S'19) received the B.S. degree in electrical engineering from the Georgia Institute of Technology, Atlanta, GA, USA, in 2013, and the M.S. degree in electrical engineering from Columbia University, New York City, NY, USA, in 2017, where he is currently pursuing the Ph.D. degree in electrical engineering. His current research interests include integrated digital low-dropout regulator designs and low-power very large scale integration circuits and systems.

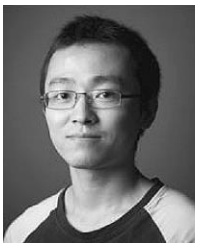

**Minhao Yang** (S'11–M'16) received the Ph.D. degree in physics from ETH Zürich in 2015. He was a Post-Doctoral Fellow with Columbia University in part supported by SNF Early Postdoc Mobility Fellowship. He is currently a scientist in the integrated circuit laboratory in EPFL. His research interests include spike coding and processing, low-power spiking sensors with embedded processing, and silicon retina and cochlea.

**Joonseong Kang**

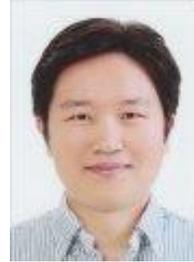

**Seungchul Jung** (S'06–M'16) received the B.S. (magna cum laude, minor: chemistry) and Ph.D. degrees in electrical engineering from the Korea Advanced Institute of Science and Technology (KAIST), Daejeon, South Korea, in 2006 and 2014, respectively.

He is currently with the Samsung Advanced Institute of Technology (SAIT), Suwon, South Korea, where he is involved in the development of wireless battery chargers and power management circuits for implantable devices. His current research interests include designing rectifiers, battery chargers, DC-DC converters, and energy harvesters for ultra-low power applications.

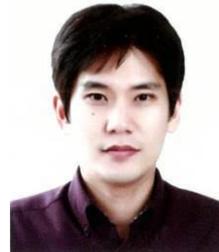

**Sang Joon Kim** received the Bachelor's degree in electrical engineering from Seoul National University, Seoul, Korea (ROK), in 2001, and the Master's degree in Applied Mathematics and the Ph.D. in Engineering Science from School of Engineering and Applied Sciences, Harvard University, Cambridge, MA, in 2007 and 2009, respectively. Since 2009, he has been with Samsung Advanced Institute of Technologies as a Principle Researcher. His current research interests include ultra low power system design, bioelectronics, neuromorphic processor, information theory and coding theory.

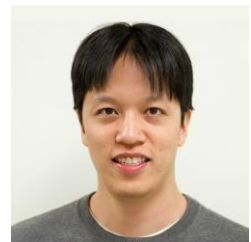

**Mingoo Seok** (S'05–M'11-SM'18) received the B.S. (summa cum laude) degree in electrical engineering from Seoul National University, Seoul, South Korea, in 2005, and the M.S. and Ph.D. degrees from the University of Michigan, Ann Arbor, MI, USA, in 2007 and 2011, respectively, all in electrical engineering. He was a member of the Technical Staff with Texas Instruments Inc., Dallas, TX, USA, in 2011. Since 2012, he has been with the Department of Electrical Engineering, Columbia University, New York, NY, USA, where he is currently an associate professor.

His current research interests include variation, voltage, aging, thermal-adaptive circuits and architecture, ultra-low power SoC design for emerging embedded systems, machine-learning VLSI architecture and circuits, and nonconventional hardware design.

Prof. Seok received the 1999 Distinguished Undergraduate Scholarship from the Korea Foundation for Advanced Studies, the 2005 Doctoral Fellowship from the Korea Foundation for Advanced Studies, and the 2008 Rackham Pre-Doctoral

Fellowship from the University of Michigan. He also received the 2009 AMD/CICC Scholarship Award for picowatt voltage reference work, and the 2009 DAC/ISSCC Design Contest for the 35-pW sensor platform design. He also received the 2015 NSF CAREER Award and the 2019 Qualcomm Faculty Award. He has been serving as an Associate Editor of the IEEE TRANSACTIONS ON CIRCUITS AND SYSTEMS—I (TCAS-I) for 2013-2015 and the IEEE TRANSACTIONS ON VERY LARGE SCALE INTEGRATION SYSTEMS (TVLSI) since 2015 and the IEEE SOLID-STATE CIRCUITS LETTER (SSCL) since 2017.